\documentclass[aps,prl,reprint,superscriptaddress,twocolumn,showkeys,amsmath,amssymb]{revtex4-1}
\usepackage[utf8]{inputenc}
\usepackage{bm}
\usepackage{graphicx}
\usepackage{amsthm}
\usepackage[english]{babel}
\usepackage{microtype}
\usepackage{mathtools}
\usepackage{dsfont}
\usepackage{mathrsfs}
\usepackage{braket}
\usepackage{hyperref}
\usepackage[shortlabels]{enumitem}
\usepackage[dvipsnames]{xcolor}
\usepackage{stmaryrd}
\usepackage{afterpage}
\usepackage{marvosym}

\hypersetup{
  colorlinks   = true, 
  urlcolor     = Blue, 
  linkcolor    = BrickRed, 
  citecolor   = PineGreen 
}

\newcommand{\id}{\openone}

\pdfpageheight\paperheight
\pdfpagewidth\paperwidth

\def\tcm{T.C.M. Group, Cavendish Laboratory, University of Cambridge, J.J. Thomson Avenue, Cambridge, CB3 0HE, UK}
\def\DAMTP{DAMTP, University of Cambridge, Wilberforce Road, Cambridge, CB3 0WA, UK}

\begin{document}

\title{SymTFT out of equilibrium: from time crystals to braided drives and Floquet codes}

\author{Vedant Motamarri}
\affiliation{\tcm}

\author{Campbell McLauchlan}
\affiliation{\DAMTP}

\author{Benjamin B\'eri}
\email[]{bfb26@cam.ac.uk}
\affiliation{\tcm}
\affiliation{\DAMTP}

\begin{abstract}
Symmetry Topological Field Theory (SymTFT) is a framework to capture universal features of quantum many-body systems by viewing them as a boundary of topological order in one higher dimension. This has yielded numerous insights in static low-energy settings. Here we study what SymTFT can reveal about nonequilibrium, focusing on one-dimensional (1D) periodically driven systems and their 2D SymTFTs. In driven settings, boundary conditions (BCs) can be dynamical and can apply both spatially and temporally. We show how this enters SymTFT via topological operators, which we then use to uncover several new results. These include revealing time crystals (TCs) as systems with symmetry-twisted temporal BCs, robust bulk ``dual TCs" in phases thought to be only boundary TCs, generating drive dualities, or identifying 2D Floquet codes as space-time duals to 1D systems with duality-twisted spatial BCs. We also show how, by making duality-twisted BCs dynamical, non-Abelian braiding of duality defects can enter SymTFT, leading to effects such as the exact pumping of symmetry charges between a system and its BCs. We illustrate our ideas for $\mathbb{Z}_2$-symmetric 1D systems, but our construction applies for any finite Abelian symmetry.
\end{abstract}

\maketitle

Recent years saw the development of deep generalizations of symmetry~\cite{Nussinov2009,nussinov2009sufficient,Cobanera2010,cobanera2011bond, KapustinSeiberg2014TQFT,GaiottoKapustinSeibergWillett2015GGS,Yoshida2016,Lake2018,mcgreevy2023generalized,Sakura_lect,bhardwaj2024lectures}. 
Among these is the insight that upon viewing a system as a boundary of topological order (TO)~\cite{kitaev2003fault,Kitaev2006,Levin-Wen,Nayak2008RMP} in one higher dimension, one can study its symmetries, their extensions to categorical symmetries involving dualities and ``dual" symmetries, and various twisted boundary conditions (BCs) in a ``theory independent" manner, i.e., insensitive to many microscopic details~\cite{Severa_2002,frohlich2004duality,frohlich2007duality,kitaev2012models,nussinov2012effective,Cobanera13,kong2017boundary,thorngren2019fusion,albert2021spin,freed2021topological,gaiotto2021orbifold,WenWei15,aasen2016topological,Aasen_cat,Apruzzi21,Kaidi22,Bhardwaj_PtII,JiWen2020categorical,lichtman2020bulk,ChatterjeeWen23,TH23,FreedMooreTeleman22,Lootens21,Lootens22}. 
The TO here is sometimes called ``symmetry topological field theory" (SymTFT)~\cite{Apruzzi21,Kaidi22,Bhardwaj_PtII,ChatterjeeWen23,TH23}. 

This uncovered unitary braided tensor categories as structures unifying topological defects describing dualities and symmetries~\cite{freed2021topological,gaiotto2021orbifold,WenWei15,aasen2016topological,Aasen_cat,Apruzzi21,Kaidi22,Bhardwaj_PtII,ChatterjeeWen23,JiWen2020categorical,lichtman2020bulk,TH23,FreedMooreTeleman22,Lootens21,Lootens22}.
The non-Abelian fusion in these categories %
render them into ``non-invertible" symmetries. 

The generality of SymTFT notwithstanding, it has thus far been used only in static  low-energy settings, yielding insights including classifications of gapped phases, dualities, or critical properties~\cite{freed2021topological,gaiotto2021orbifold,WenWei15,aasen2016topological,Aasen_cat,Apruzzi21,Kaidi22,Bhardwaj_PtII,ChatterjeeWen23,JiWen2020categorical,lichtman2020bulk,TH23,FreedMooreTeleman22,Lootens21,Lootens22,Lootens23,BhardwajcatLand,Bhardwaj1_1,Pace23,ChatterjeeWen23_gapless,WenPotter_SPTgapless,Huang_Chen_criticality,bhardwaj2023club}. 

What can SymTFT reveal about nonequilibrium? 
Here we study this, focusing on one-dimensional (1D) Floquet systems and their 2D SymTFTs.
We take the former many-body localized (MBL)~\cite{Fleish80,Gornyi2005,basko2006metal,Huse2013LPQO,serbyn2013local,Huse_MBL_phenom_14,chandran2015constructing,ros2015integrals,Wahl2017PRX,Goihl2018,NandkishoreHuse_review,AltmanReview,Abanin2017,Alet2018,ImbrieLIOMreview2017}; this allows for genuinely dynamical phases, e.g., time crystals (TCs)~\cite{Wilczek2012quantumTC,ShapereWilczek2012classicalTC,WatanabeOshikawa2015TC,KhemaniLazaridesMoessnerSondhi2016TC,ElseTC2016,KeyserlingkSondhi2016floquetSPT,KeyserlingkSondhi2016floquetSSB,KeyserlingkKhemaniSondhi2016stability,KhemaniKeyserlingkSondhi2017RepTh,time_crystals_review,Sacha2018,Else2020}. 
Driven settings can involve both spatial and temporal BCs, and BCs can be dynamical. 
As we shall show, this adds novel features to SymTFT which, in turn, yield new insights in 1D %
that reveal, e.g., TCs as featuring symmetry-twisted temporal BCs, %
bulk ``dual TCs" in phases thought to be only boundary TCs, or dynamical dualities. 
We also show how space-time duality of BCs, when applied to duality-twisted BCs, yields 2D Floquet codes (FCs)~\cite{Hastings2021dynamically,Vuillot21,DavydovaPRXQ23,kesselring2022anyon,vu2023measurementinduced,sullivan2023floquet,aasen2023measurement,ellison2023floquet,davydova2023quantum,dua2023engineering}, thus unifying TCs and FCs via SymTFT.
Furthermore we show that,
by making duality-twisted BCs dynamical, non-Abelian braiding---not only 
fusion---of duality defects~\cite{Bombin10,BarkeshliQi_PRX12,YouWen12,Barkeshli13c,BBCW19}
can enter SymTFT, with remarkable consequences such as the exact pumping of symmetry charges between a system and its BCs. 

In what follows we assume that the 1D system has a finite Abelian global symmetry $G$. 
Since we find striking new results already for $G=\mathbb{Z}_2$, we shall use this simplest case to illustrate our ideas. 
We may take the 2D bulk as fictitious, but our construction also extends to boundaries of a physical 2D TO, including TO TCs~\cite{KeyserlingkKhemaniSondhi2016stability,KhemaniKeyserlingkSondhi2017RepTh,wahl2023topologically}.

\begin{figure}[t]
 \includegraphics[scale=0.75]{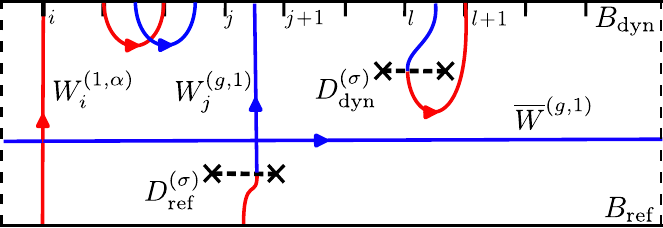}
 \caption{Embedding the 1D system as boundary $B_\text{dyn}$ of a 2D TO on a cylinder (dashed sides are identified). Boundary $B_\text{ref}$ condenses $(1,\alpha)$ anyons. $W^{(1,\alpha)}$ lines are shown in red, $W^{(g,1)}$ in blue. The $W_i^{(1,\alpha)}$  serve as local order parameters in 1D; $\smash{\overline{W}}^{(g,1)}$  are global symmetries. Also shown are $W^{(a)}_{j-2,j-1}$ for $a\in\{(1,\alpha),(g,1)\}$, as is $W^{(1,\alpha)}_{l,l+1}$, crossing domain wall $D_\text{dyn}^{(\sigma)}$.
 $D_\text{ref}^{(\sigma)}$ enables $g$-twisting boundary conditions via %
 $W_{j}^{(g,1)}$ (ending on link $\langle j,j+1\rangle$)~\cite{Wdetails}. Crosses represent twist defects.}\label{fig:setup}
\end{figure}

\textit{Hilbert space, symmetries, order parameters:} We first summarize 
the (condensed matter) SymTFT construction,
bridging between Refs.~\onlinecite{JiWen2020categorical,lichtman2020bulk,ChatterjeeWen23,TH23} by our approach to order parameters, degeneracies, and BCs~\cite{SM}.
We focus on 1D spin chains with closed BCs and $L$-site Hilbert space $\mathcal{H}_{\text{dyn}}=\mathcal{H}_{1}^{\otimes L}$; for $G=\mathbb{Z}_{2}$, $\mathcal{H}_{1}=\text{span}(\{\ket 0,\ket1\})$~\cite{SM}.
The global symmetry acts as $V_{g}=\prod_{k=1}^{L}v_{g}^{(k)}$, $g\in G$, with onsite unitary $v_{g}^{(k)}$ at site $k$. The ``patch" symmetry~\cite{JiWen2020categorical}, for interval (``patch'') $[i,j]$ is $V_{g}^{(ij)}=\prod_{k=i}^{j}v_{g}^{(k)}$.
We view the 1D system as being on a ``dynamical'' boundary $B_{\text{dyn}}$ of a 2D TO, cf.~Fig.~\ref{fig:setup}. 
The TO is $G$-TO [i.e., quantum double $\mathcal{D}(G)$ or $G$-gauge theory~\cite{kitaev2003fault,Kitaev2006,Levin-Wen}] with Abelian ``flux-charge composite" anyons $a=(g,\alpha)$ with ``flux'' $g\in G$ and ``charge'' $\alpha\in\text{Rep}\,G$ [the set of irreducible representations $\chi_{\alpha}:G\to U(1)$ of $G$]. 
Fusing $a$ with $b=(g',\alpha')$ yields $ab=(gg',\alpha\alpha')$, where $\chi_{\alpha\alpha'}=\chi_{\alpha}\chi_{\alpha'}$; the $a$-antianyon is thus $\bar{a}=(g^{-1},\alpha^{-1})$. Encircling $a$ with $b$ yields braiding phase $e^{i2\theta_{ab}}=\chi_{\alpha}(g')\chi_{\alpha'}(g)$, while exchanging two $a$ accrues $e^{i\theta_{aa}}=\chi_{\alpha}(g)$. For $G=\mathbb{Z}_{2}$, the anyons are $\bm{1}=(1,1)$, $e=(1,-1)$, $m=(-1,1)$ and $f=em$, with $e^{i2\theta_{em}}=e^{i\theta_{ff}}=-1$, $e^{i\theta_{ee}}=e^{i\theta_{mm}}=1$.
The $G$-TO has gappable (and many-body localizable) edges; a gapped (MBL) edge
condenses a ``Lagrangian subgroup'' $\mathcal{A}$: a maximal subset of self-bosons that braid trivially with each other~\cite{Kapustin_2011,Levin_2013,Barkeshli13b}. 

For the 2D setup to reduce to 1D, we require that away from $B_{\text{dyn}}$ there be a gap much larger than $B_{\text{dyn}}$ energy scales. 
Thus, we identify $\mathcal{H}_{\text{dyn}}$ as the subspace with zero anyons away from $B_{\text{dyn}}$. 
The 1D  algebra is then the ``boundary algebra'' (BA): that of the TO's unitary string operators $W^{(a)}$ (more precisely, their equivalence classes under endpoint-preserving deformations) that may create anyons $a$ at and only at $B_{\text{dyn}}$. %
Anyons are the endpoints of $W^{(a)}$. Thus, the BA includes open strings $W_{ij}^{(a)}$ ending on an $a$-$\bar{a}$ pair at $i,j\in B_{\text{dyn}}$, cf. Fig.~\ref{fig:setup}~\cite{Wdetails}. 
By the TO's fusion and braiding properties, respectively~\cite{JiWen2020categorical,lichtman2020bulk,ChatterjeeWen23,TH23}, %
\begin{equation}
\!\!W_{ij}^{(a)}W_{ij}^{(b)}=W_{ij}^{(ab)}\!\!,\ W_{ij}^{(b)}W_{kl}^{(a)}=e^{i2\eta_{ij,kl}\theta_{ab}}W_{kl}^{(a)}W_{ij}^{(b)}\!\!,\label{eq:cat_symm}
\end{equation}
with $\eta_{ij,kl}=\pm1$ (depending on orientations) if the strings intersect an odd number of times and $\eta_{ij,kl}=0$ otherwise. 

To relate Eq.~\eqref{eq:cat_symm} to the 1D algebra, we first identify $W_{i-1,j}^{(g,1)}\equiv V_{g}^{(ij)}$~\cite{Wdetails}. %
For $G=\mathbb{Z}_{2}$ this is $W_{i-1,j}^{(m)}=\prod_{i\leq k\leq j}X_{k}$, with $X_{k}$, $Y_{k}$, $Z_{k}$ the Pauli operators on site $k$. 
The global symmetry is $\smash{\overline{W}}^{(g,1)}=V_{g}$, where $\smash{\overline{W}}^{(g,1)}$ encircles $B_{\text{dyn}}$, cf. Fig.~\ref{fig:setup}.  For $G=\mathbb{Z}_{2}$,  $\smash{\overline{W}}^{(m)}=\prod_{k=1}^L X_{k}$.

Next, we implement local (on-site) order parameters $O_{k}^{(\alpha)}$, obeying $V_{g}^{(ij)}O_{k}^{(\alpha)}V_{g}^{(ij)\dagger}=\chi_{\alpha}(g)O_{k}^{(\alpha)}$ for $i\leq k\leq j$. 
By $W_{i-1,j}^{(g,1)}=V_{g}^{(ij)}$, 
we get the same relation if $O_{k}^{(\alpha)}=W_{k}^{(1,\alpha)}$, a string with a \emph{single} endpoint $k\in B_{\text{dyn}}$. 
To ensure that the other endpoint creates no anyons, we place the 2D TO on a cylinder and emanate $W_{k}^{(1,\alpha)}$ from the other, gapped, ``reference'' boundary $B_{\text{ref}}$, with Lagrangian subgroup   $\mathcal{A}_{\text{ref}}=\langle\{(1,\alpha)|\alpha\in\text{Rep}\,G\}\rangle$, cf. Fig.~\ref{fig:setup}.
For $G=\mathbb{Z}_{2}$, we have $\mathcal{A}_{\text{ref}}=\langle e\rangle$, $W_{k}^{(e)}\equiv Z_{k}$.

While $W_{ij}^{(g,1)}$ and $W_{k}^{(1,\alpha)}$ express standard 1D concepts, we must comment on their ``electric-magnetic dual" $W_{ij}^{(1,\alpha)}$ and $W_{k}^{(g,1)}$~\cite{Wdetails}.
The $W_{ij}^{(1,\alpha)}$ describe the ``dual symmetry" $\tilde{G}$, the one broken in a \emph{disordered} phase~\cite{JiWen2020categorical}. 
By incorporating both $G$ and $\tilde{G}$, Eq. \eqref{eq:cat_symm} expresses the  categorical symmetries in 1D~\cite{JiWen2020categorical}.
For $G=\mathbb{Z}_{2}$, we denote $\tilde{G}=\tilde{\mathbb{Z}}_{2}$, and by $W_{i}^{(1,\alpha)\dagger}W_{j}^{(1,\alpha)}=W_{ij}^{(1,\alpha)}$ (via deformations), we have $W_{ij}^{(e)}=Z_iZ_j$.
For the ``dual order parameter" $W_k^{(g,1)}$, we must discuss twisted BCs. 
This is what we do next, via ``duality defects" on the cylinder.

\textit{Twisted spatial BCs:}  TO comes with anyonic symmetries $\sigma$~\cite{Kitaev2006,Bombin10,BarkeshliQi_PRX12,YouWen12,Barkeshli13c,BBCW19}: automorphisms $a'=\sigma(a)$ preserving fusion and braiding properties. 
They may furnish domain walls $D^{(\sigma)}$ such that
string $W^{(a)}$ must continue as $W^{[\sigma(a)]}$ across $D^{(\sigma)}$  to avoid creating an anyon at $D^{(\sigma)}$, cf. Fig.~\ref{fig:setup}. 
A finite-length $D^{(\sigma)}$ ends on ``twist defects''~\cite{Bombin10,BarkeshliQi_PRX12,YouWen12,Barkeshli13c,BBCW19}. %
Automorphisms include %
electric-magnetic dualities, $\sigma(g)\in\text{Rep}\,G$, $\forall g\in G$, $\sigma^{2}=1$. 
We mostly focus on these henceforth. 
For $G=\mathbb{Z}_{2}$ this is $\sigma(e)=m$. 

With $D^{(\sigma)}$ we now $g$-twist BCs. (We return to duality-twisting later.) 
To this end, we add a short wall $D_{\text{ref}}^{(\sigma)}$ near $B_{\text{ref}}$, cf. Fig.~\ref{fig:setup}; crossing this, $(1,\alpha)$ emanating at $B_{\text{\text{ref}}}$ yields $W_j^{(g,1)}$ with $g=\sigma(\alpha)$, i.e., an unpaired $(g,1)$ anyon at 
$B_{\text{dyn}}$~\cite{Wdetails}. 
This $g$-twists BCs via the $G$-Aharonov-Bohm effect: 
$\smash{\overline{W}}^{(1,\alpha)}$, bringing a charge $(1,\alpha)$ around the system, detects the flux $g$  by $\smash{\overline{W}}^{(1,\alpha)}\!\ket{\psi'}=\chi_\alpha(g)\ket{\psi'}$, where $\ket{\psi'}=W_{j}^{(g,1)}\!\ket{\psi}$ and $\smash{\overline{W}}^{(1,\alpha)}\ket{\psi }=\ket{\psi}$~\cite{SM}. 

For $G=\mathbb{Z}_{2}$, %
$W_{j}^{(m)}W_{k,k+1}^{(e)}W_{j}^{(m)\dagger}=(-1)^{\delta_{jk}}W_{k,k+1}^{(e)}$; thus $W_{j}^{(m)}$ flips $Z_{k}Z_{k+1}$ only for $k=j$. %
No $X_k$ product  %
does this.
Thus, %
we gauge %
via
$Z_{j}Z_{j+1}\to Z_{j}\tilde{Z}_{j}Z_{j+1}\equiv W_{j,j+1}^{(e)}$, yielding $W_{j}^{(m)}\equiv\tilde{X}_{j}$,  with link Paulis $\tilde{Z}_{j}$, $\tilde{X}_{j}$. 
In a gauge with just one dynamic link, %
i.e., $W_{k,k+1}^{(e)}=Z_{k}Z_{k+1}$ ($k\neq j$), we have $\smash{\overline{W}}^{(e)}=\prod_{j=1}^L W_{j,j+1}^{(e)}=\tilde{Z}_{j}$ detecting untwisted (periodic) and $m$-twisted (antiperiodic) BCs~\cite{SM}.

\textit{1D Floquet drives:} We now turn to dynamics. We shall mostly focus on 1D Floquet unitaries
of the form 
\begin{equation}
U_{\text{F}}^{(b,\mathcal{A})}=\smash{\overline{W}}^{(b)}e^{-iH_{1}}e^{-iH_{0}^{(\mathcal{A})}}.
\label{eq:UFbA}
\end{equation}
Here $H_{0}^{(\mathcal{A})}$ is a local $G$- (and $\tilde{G}$-) symmetric Hamiltonian with mutually commuting terms; it sets the dominant energy scale thus its Lagrangian subgroup $\mathcal{A}$ sets the eigenstate order~\cite{Huse2013LPQO}. 
We take $H_{0}^{(\mathcal{A})}$ to be disordered to ensure MBL~\cite{Fleish80,Gornyi2005,basko2006metal,Huse2013LPQO,serbyn2013local,Huse_MBL_phenom_14,chandran2015constructing,ros2015integrals,Wahl2017PRX,Goihl2018,NandkishoreHuse_review,AltmanReview,Abanin2017,Alet2018,ImbrieLIOMreview2017}.
For $G=\mathbb{Z}_{2}$, we take $H_{0}^{(\langle a\rangle)}=\sum_{j}J_{j}W_{j,j+1}^{(a)}+J_{j}^{(\text{int})}W_{j,j+2}^{(a)}$ (with $0<J_{j}<\pi/2$ random with typical value $\bar{J}\sim\pi/4$ and $|J_{j}^{(\text{int})}|\ll\bar{J}$). 
This gives paramagnet (PM) eigenstates for $a=m$ [recall $W_{j-1,j}^{(m)}=X_{j}$] and spin-glass (SG) for $a=e$~\cite{Huse2013LPQO,KhemaniLazaridesMoessnerSondhi2016TC,ElseTC2016,KeyserlingkSondhi2016floquetSPT,KeyserlingkSondhi2016floquetSSB,KeyserlingkKhemaniSondhi2016stability,KhemaniKeyserlingkSondhi2017RepTh}.
$H_{1}$ adds local $G$- (and $\tilde{G}$-) symmetric perturbations. 
For $G=\mathbb{Z}_{2}$, $H_{1}=\sum_{j}h_{j}W_{j,j+1}^{(a')}+h_{j}^{(\text{int})}W_{j,j+2}^{(a')}$ with $a'=\sigma(a)$
and $|h_{j}|,|h_{j}^{(\text{int})}|\ll\bar{J}$. 
Since $\smash{\overline{W}}^{(b)}$ commutes with all terms in $H_{0,1}$, it does not affect eigenstates.
However, for $b\notin\mathcal{A}$, as we next show, we get TCs.

\textit{TC SymTFT:}
We first illustrate this for the paradigmatic $\mathbb{Z}_{2}$ TC, the $\pi$SG~\cite{KhemaniLazaridesMoessnerSondhi2016TC,ElseTC2016,KeyserlingkSondhi2016floquetSPT,KeyserlingkSondhi2016floquetSSB,KeyserlingkKhemaniSondhi2016stability,KhemaniKeyserlingkSondhi2017RepTh,time_crystals_review,Sacha2018,Else2020}. 
This arises for $b=m$ and $\mathcal{A}=\langle e\rangle$. 
We start with the zero-correlation-length (or fixed-point, FP) limit, $H_{1}=0$. 
Here $W_{k,k+1}^{(e)}$ %
are local integrals of motion (LIOMs)~\cite{serbyn2013local,Huse_MBL_phenom_14,chandran2015constructing,ros2015integrals,Wahl2017PRX,Goihl2018,NandkishoreHuse_review,AltmanReview,Abanin2017,Alet2018,ImbrieLIOMreview2017} and the eigenstates of $U_{\text{F}}^{(m,\langle e\rangle)}$ are $\ket n=\ket{\{d_{k}\},p}$, with $d_{k},p\in\{-1,1\}$ where $W_{k,k+1}^{(e)}\ket n=d_{k}\ket n$, $\smash{\overline{W}}^{(m)}\ket n=p\ket n$. 
These are SG eigenstates, displaying long-ranged order-parameter correlations 
$C_{n;ij}=\bra nW_{ij}^{(e)}\ket n \neq 0$.
(We henceforth write $C_{n;ij}=C_{ij}$ for brevity.)

Furthermore, by $\smash{\overline{W}}^{(m)\dagger}W_{i}^{(e)}\smash{\overline{W}}^{(m)}=e^{-2i\theta_{em}}W_{i}^{(e)}$
and $H_{0}^{(\langle e\rangle)}W_{i}^{(e)}=W_{i}^{(e)}H_{0}^{(\langle e\rangle)}$, at the FP we have $W_{i}^{(e)}(t)=[U_{\text{F}}^{(m,\langle e\rangle)\dagger}]^{t}W_{i}^{(e)}[U_{\text{F}}^{(m,\langle e\rangle)}]^{t}=e^{-2it\theta_{em}}W_{i}^{(e)}$
($t\in\mathbb{Z}$) hence
$C_{ij}(t)=\bra nW_{i}^{(e)}(t)W_{j}^{(e)}\ket n=(-1)^{t}C_{ij}$.
This period doubled long-ranged order is the spatio-temporal order defining TCs~\cite{KeyserlingkKhemaniSondhi2016stability,KhemaniKeyserlingkSondhi2017RepTh,time_crystals_review}.
It also comes with spectral multiplets: if $U_{\text{F}}^{(m,\langle e\rangle)}\ket n=e^{-i\varepsilon_{n}}\ket n$,
then, at the FP, $U_{\text{F}}^{(m,\langle e\rangle)}W_{i}^{(e)}\ket n=e^{2i\theta_{em}}e^{-i\varepsilon_{n}}W_{i}^{(e)}\ket n$;
by $2\theta_{em}=\pi$, this is the familiar $\pi$ spectral pairing~\cite{KeyserlingkKhemaniSondhi2016stability,KhemaniKeyserlingkSondhi2017RepTh,time_crystals_review}.

Away from the FP, for weak $H_{1}\neq0$, similar considerations apply, but now $e^{-iH_{1}}e^{-iH_{0}^{(\mathcal{A})}}$ has LIOMs $\mathcal{W}_{k,k+1}^{(a)}=U_{\text{LU}}W_{k,k+1}^{(a)}U_{\text{LU}}^{\dagger}$ ($a\in\mathcal{A}$) with $U_{\text{LU}}$ a ($H_{1}$-dependent) local unitary~\cite{serbyn2013local,Huse_MBL_phenom_14,chandran2015constructing,ros2015integrals,Wahl2017PRX,Goihl2018,NandkishoreHuse_review,AltmanReview,Abanin2017,Alet2018,ImbrieLIOMreview2017} (mapping local to quasi-local operators with tails decaying
exponentially on the scale of the localization length $\xi$). 
For a $G$-symmetric system $U_{\text{LU}}\smash{\overline{W}}^{(b)}=\smash{\overline{W}}^{(b)}U_{\text{LU}}$ and thus the eigenstates are now $\ket n=U_{\text{LU}}\ket{\{d_{k}\},p}$.
In the SymTFT, we embed $U_{\text{LU}}$ such that it acts as $\id$ away from $B_{\text{dyn}}$. Thus, all the previous considerations apply upon suitably conjugating by $U_{\text{LU}}$, with the same results up to corrections $\sim e^{-|i-j|/\xi}$ for correlations and $\sim e^{-L/\xi}$ for spectral pairing~\cite{KeyserlingkKhemaniSondhi2016stability,KhemaniKeyserlingkSondhi2017RepTh,time_crystals_review}.

\textit{Dual TC order:} The above discussion holds beyond $G=\mathbb{Z}_{2}$. %
However, $G=\mathbb{Z}_{2}$ already offers striking new results. 
First we note that $W_{i}^{(e)}$ and $W_{i}^{(m)}$ differ only by electric-magnetic duality (cf.~Fig.~\ref{fig:setup}). 
This implies \emph{bulk} ``dual TC'' features for $U_{\text{F}}^{(e,\langle m\rangle)}$,
i.e., the ``$0\pi$PM" thus far thought to be only a boundary TC~\cite{Wilczek2012quantumTC,ShapereWilczek2012classicalTC,WatanabeOshikawa2015TC,KhemaniLazaridesMoessnerSondhi2016TC,ElseTC2016,KeyserlingkSondhi2016floquetSPT,KeyserlingkSondhi2016floquetSSB,KeyserlingkKhemaniSondhi2016stability,KhemaniKeyserlingkSondhi2017RepTh,time_crystals_review,Sacha2018,Else2020}. 

\begin{figure}[t]
 \includegraphics[width=\columnwidth]{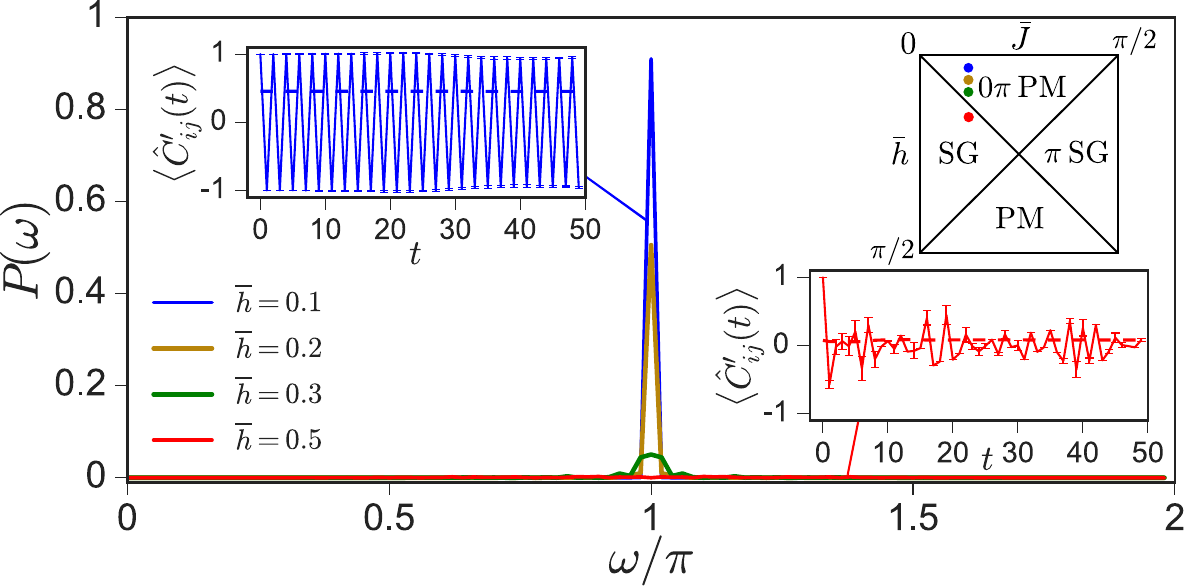}
 \caption{Dual TC correlations in time and frequency domain. 
 The time series show $\langle\hat{C}_{ij}^{\prime}(t)\rangle=\langle C_{ij}^{\prime}(t)/C_{ij}^{\prime}\rangle$, with $\langle |C_{ij}^{\prime}|\rangle$ dashed and $|i-j|=L/2$, where $\langle\ldots\rangle$ is the average over eigenstates and $1000$ disorder realizations~\cite{h05fn}. 
 $P(\omega)$ is the power spectrum of $\langle\hat{C}_{ij}^{\prime}(t)\rangle$. 
 We used $U_{\text{F}}^{(e,\langle m\rangle)}$ with $L=10$, sampled $h_j$ uniformly from $[\bar{h}/2, 3\bar{h}/2]$, and $J_j$ from $[\bar{J}/2,3\bar{J}/2]$, with $\bar{J}=\pi/8$. 
 For $\bar{h}=0.5$ the system is in the static SG phase, as seen from absorbing $\smash{\overline{W}}^{(e)}$ via the $W^{(e)}_{j,j+1}$ terms in $e^{-iH_1}$.
  }\label{fig:dualTC_data}
\end{figure}

We study these via $C_{ij}^{\prime}(t)=\bra nW_{i}^{(m)}(t)W_{j}^{(m)}\ket n$,
where $\ket n=\ket{\{x_{k}\}}$, $X_{k}\ket n=x_{k}\ket n$ ($x_{k}\in\{-1,1\}$)
are PM eigenstates. (For brevity, $X_{k}$ also denotes the LIOMs away from the FP.) 
By $W_{i}^{(m)}(t)$, BCs became dynamic. 
To place $C_{ij}^{\prime}(t)$ in $\mathcal{H}_\text{dyn}$, we eliminate link variables via~\cite{Wdetails}
\begin{equation}
W_{i}^{(m)}(t)W_{j}^{(m)}=[U_{\text{F}}^{(e,\langle m\rangle)\dagger}]^{t}W_{ij}^{(m)}[U_{\text{F}}^{(e,\langle m\rangle)\prime}]^{t},
\end{equation}
where $U_{\text{F}}^{(e,\langle m\rangle)}\!\!\to U_{\text{F}}^{(e,\langle m\rangle)\prime}$ takes $W_{k,k+1}^{(e)}\!\!\to(-1)^{\delta_{jk}}W_{k,k+1}^{(e)}$, hence 
$h_{j}\to-h_{j}$, $h_{k}^{(\text{int})}\!\!\!\to-h_{k}^{(\text{int})}$
($k=j-1,j$), and $\smash{\overline{W}}^{(e)}\!\!\to-\smash{\overline{W}}^{(e)}$. 
For $|i-j|\gg \xi$, we have $C_{ij}^{\prime}(t)=(-1)^{t}C_{ij}^{\prime}$ with $C_{ij}^{\prime}=\bra n W_{ij}^{(m)}\ket n$. 
This is long-ranged in PM eigenstates~\cite{fradkin2017disorder} (signaling $\tilde{\mathbb{Z}}_{2}$ breaking~\cite{JiWen2020categorical}), thus the period-doubled $C_{ij}^{\prime}(t)$ signals spatio-temporal $\tilde{\mathbb{Z}}_{2}$ TC order. 
We illustrate $C_{ij}^{\prime}(t)$ numerically in Fig.~\ref{fig:dualTC_data}. 

The $(-1)^t$ in $C_{ij}^{\prime}(t)$ here also links to $\pi$ spectral pairing, now between $\ket n$ and $\smash{W_{i}^{(m)}}\ket n$, i.e., between \emph{distinct} BCs, or, within $\mathcal{H}_\text{dyn}$, between $\smash{U_{\text{F}}^{(e,\langle m\rangle)}}$ and $\smash{U_{\text{F}}^{(e,\langle m\rangle)\prime}}$ spectra with the same $\smash{\overline{W}}^{(m)}$ eigenvalue. 
This is the dual TC counterpart of the inter-BC degeneracy of a static PM. 

The $\tilde{\mathbb{Z}}_{2}$ TC also inherits the $\pi$SG's ``absolute stability'':  robustness against $\mathbb{Z}_{2}$-breaking perturbations~\cite{KeyserlingkKhemaniSondhi2016stability}. 
In the $\pi$SG, this arises from MBL and $\pi$ pairing, the latter (by lifting degeneracy) excluding symmetry-broken unperturbed eigenstates. 
The same holds for the $\tilde{\mathbb{Z}}_{2}$ TC. 
However, it also does for the static PM; the less trivial fact is that $\tilde{\mathbb{Z}}_{2}$ TCs are also absolutely stable against $\tilde{\mathbb{Z}}_{2}$ breaking.
We numerically illustrate this in Ref.~\onlinecite{SM}.

\textit{Drive dualities:} The (dual) TC drives with $b\notin\mathcal{A}$ in Eq.~\eqref{eq:UFbA} are complemented by the static $\mathcal{A}=\langle b\rangle$ drives; for $\mathbb{Z}_2$, these together yield $\pi$SG, $0\pi$PM, SG and PM as the four Floquet phases. 
The SymTFT supplies drive dualities: automorphisms $\sigma,\sigma': \mathcal{A}\to \sigma(\mathcal{A}), b\to \sigma'(b)$. 
Here $\sigma$ generates ``static" (Kramers-Wannier-type) dualities, 
while $\sigma'$ are ``Floquet dualities"~\cite{Berdanier_2018}. 
Within SymTFT, the former corresponds to a domain wall $D^{(\sigma)}_L$ wrapping $B_\text{dyn}$~\cite{freed2021topological,gaiotto2021orbifold,WenWei15,aasen2016topological,Aasen_cat,Apruzzi21,Kaidi22,Bhardwaj_PtII,ChatterjeeWen23,JiWen2020categorical,lichtman2020bulk,TH23,FreedMooreTeleman22,Lootens21,Lootens22,Lootens23}, yielding a topological, timelike, worldsheet in spacetime. 
The latter is also implemented topologically, via a domain wall worldsheet  $D^{(\sigma^\prime)}_\circ$ embracing $\smash{\overline{W}}^{(b)}$.

\begin{figure}[t]
 \includegraphics[width=0.95\columnwidth]{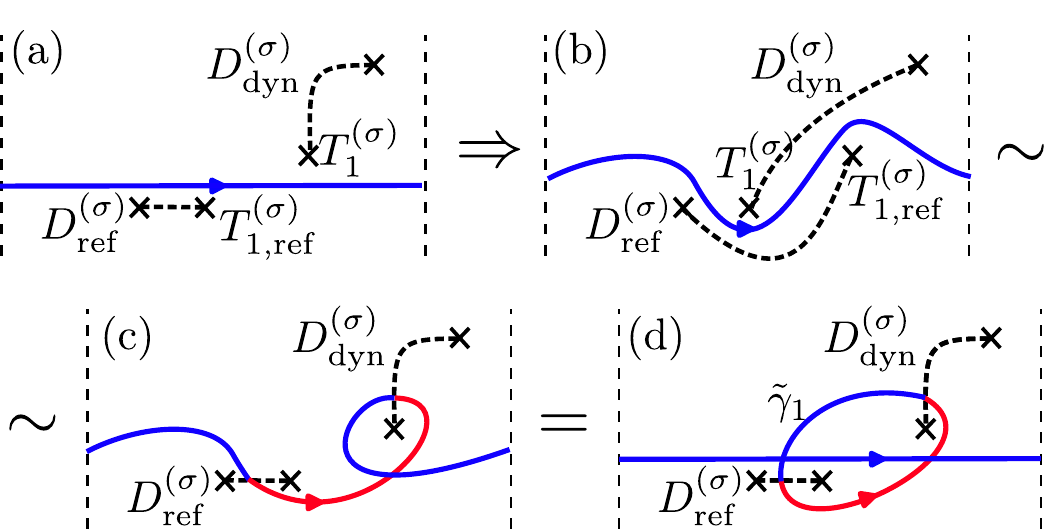}
 \caption{Braid behind the braided drive $U_{\text{F},R}^{(b,\mathcal{A}),\sigma}$ and the corresponding evolution of $\smash{\overline{W}}^{(b)}$ (blue line) through deformations avoiding the twist defects (crosses). The two sides of $\sim$ are related by a contractible loop encircling $T_{1,\text{ref}}^{(\sigma)}$. The sequence shows that $\smash{\overline{W}}^{(b)}$ evolves into $\tilde{\gamma}_{1} \smash{\overline{W}}^{(b)}$ (up to a phase), where the exact zero mode $\tilde{\gamma}_{1}$ corresponds to the noncontractible blue-red loop crossing both $D_{\text{dyn}}^{(\sigma)}$ and $D_{\text{ref}}^{(\sigma)}$.
  }\label{fig:braid}
\end{figure}

\textit{Duality-twisted BCs and braided drives:} Next, we discuss duality-twisted BCs~\cite{schutz1993duality,Grimm_2002,WenWei15,aasen2016topological,Aasen_cat,tan2022topological,Mitra23,samanta2023isolated}
and how, in driven settings, these may lead to non-Abelian defect braiding. 
We focus on $G=\mathbb{Z}_{2}$, but the generalization will be clear.  
Consider a short domain wall $D_{\text{dyn}}^{(\sigma)}$ along $B_{\text{dyn}}$, from $l\!-\!1/4$ [with twist defect $T_{1}^{(\sigma)}$] to $l\!+\!3/4$ [with $T_{2}^{(\sigma)}$],  cf. Fig.~\ref{fig:setup}.
We define the BA such that its members ending on $l$ or $\langle l,l\!+\!1\rangle$ %
cross $D_{\text{dyn}}^{(\sigma)}$. %
Now we push $T_{1}^{(\sigma)}$ into the 2D bulk, cf.~Fig.~\ref{fig:braid}a. 
This pushes $W_{l-1,l}^{(e)}$ and $W_{l-1,l}^{(m)}$ far from $B_{\text{dyn}}$~\cite{Wdetails}; %
thus rendering $Z_{l-1}Z_{l}$ and $X_{l}$ \emph{nonlocal}. 
Their product, $W_{l-1,l}^{(f)}=Z_{l-1}Y_{l}$, is however local since $\sigma(f)=f$ allows us to pull $W_{l-1,l}^{(f)}$ through $T_{1}^{(\sigma)}$ to $B_{\text{dyn}}$.

To duality twist BCs, we allow only local BA members in Hamiltonians, including 
in $H_{0}^{(\mathcal{A})}$ and $H_{1}$. 
To our knowledge, this locality perspective, afforded by SymTFT, is a novel view on duality-twisted BCs~\cite{dualtwist}. 

To complete duality-twisting $U_{\text{F}}^{(b,\mathcal{A})}$, we view $\smash{\overline{W}}^{(b)}$ as a topological ingredient; thus, although it is \emph{not} a product of local BA members for $b\in\{e,m\}$ [by featuring $W_{l-1,l}^{(b)}$], we retain it in the duality-twisted drive, denoted by $U_{\text{F}}^{(b,\mathcal{A}),\sigma}$.

The first striking feature is that $U_{\text{F}}^{(b,\mathcal{A}),\sigma}$ and its $f$-twisted pair $U_{\text{F}}^{(b,\mathcal{A}),\sigma,f}\!\!\!=\!W^{(f)\dagger}_l U_{\text{F}}^{(b,\mathcal{A}),\sigma}W^{(f)}_l $ have \emph{exactly} $\pi$-paired spectra. 
This is by $W_{l}^{(f)}= W_{l-1}^{(m)}W_{l}^{(e)\dagger}=\tilde{X}_{l-1}Z_{l}$
commuting with \emph{all} local BA members (upon gauging $Z_{l-1}Y_{l}\to Z_{l-1}\tilde{Z}_{l-1}Y_{l}$), and thus with $H_{0}^{(\mathcal{A})}$ and $H_{1}$, while anticommuting with $\smash{\overline{W}}^{(b)}$ ($b\in\{e,m\}$).

In $W_{l}^{(f)}$ we recognize a loop encircling $T_{1}^{(\sigma)}$ and a twist from $D_{\text{ref}}^{(\sigma)}$, i.e., a twist-quartet logical operator of the TO as a quantum code, cf.~Fig.~\ref{fig:braid}d.
By $U_{\text{F}}^{(b,\mathcal{A}),\sigma\dagger}W_{l}^{(f)}U_{\text{F}}^{(b,\mathcal{A}),\sigma}=-W_{l}^{(f)}$ and the localization to $l$ (in terms of the 1D system), we also recognize $W_{l}^{(f)}\equiv\tilde{\gamma}_{1}$ as a $\pi$ mode (generalizing $\pi$-Majorana fermions of TCs)---that is now exact. 
As in other TCs, this $\pi$ mode has a pair $\tilde{\gamma}_{2}$, with which $i\tilde{\gamma}_{2}\tilde{\gamma}_{1}=\mathcal{W}_{\sigma}^{(a)}$ is a nonlocal integral of motion. 
Here, $a\text{\ensuremath{\in\mathcal{A}}}$ and $\mathcal{W}_{\sigma}^{(a)}=U_{\text{LU}}W_{\sigma}^{(a)}U_{\text{LU}}^{\dagger}$ with $W_{\sigma}^{(a)}$ the (BA deformation of) a noncontractible loop encircling $D_{\text{dyn}}^{(\sigma)}$ [i.e., $W_{\sigma}^{(a)}$ is the conjugate twist-quartet logical]. 
By $\tilde{\gamma}_{1}\mathcal{W}_{\sigma}^{(a)}=e^{2i\theta_{fa}}\mathcal{W}_{\sigma}^{(a)}\tilde{\gamma}_{1}=-\mathcal{W}_{\sigma}^{(a)}\tilde{\gamma}_{1}$, we have $\tilde{\gamma}_{1}\tilde{\gamma}_{2}=-\tilde{\gamma}_{2}\tilde{\gamma}_{1}$, as befits fermions, and from $\mathcal{W}_{\sigma}^{(a)}U_{\text{F}}^{(b,\mathcal{A}),\sigma}=U_{\text{F}}^{(b,\mathcal{A}),\sigma}\mathcal{W}_{\sigma}^{(a)}$ we find that $\tilde{\gamma}_{2}$ is another exact $\pi$ mode. 
We stress that $\tilde{\gamma}_{1,2}$, and the corresponding spectral multiplets, are present exactly for any $G$-symmetric local $U_{\text{F}}^{(b,\mathcal{A}),\sigma}$ even without strong MBL~\cite{SM} and they even survive local $G$-breaking perturbations.

Next, we discuss how making duality-twisting dynamic leads to drives featuring non-Abelian defect braids. 
We take $D_{\text{dyn}}^{(\sigma)}$ to be present throughout, and denote by $\tilde{D}_{l}^{(\sigma)}$ the act of pushing $T_{1}^{(\sigma)}$ into the 2D bulk (thus duality-twisting BCs). 
Braids enter via the interplay of $\tilde{D}_{l}^{(\sigma)}$ with $D_{\text{ref}}^{(\sigma)}$; for concreteness we focus on a single exchange $R$ between  $T_{1}^{(\sigma)}$ and one of the $D_{\text{ref}}^{(\sigma)}$ twists, cf. Fig.~\ref{fig:braid}. %
We denote this braided move by  $\tilde{D}_{l}^{(\sigma,R)}=R\tilde{D}_{l}^{(\sigma)}$. 

Introducing $[\tilde{D}_{l}^{(\sigma,R)}](t)=[U_{\text{F}}^{(b,\mathcal{A})\dagger}]^t[\tilde{D}_{l}^{(\sigma,R)}][U_{\text{F}}^{(b,\mathcal{A})}]^t$ we make the braided twist dynamic, which we study via  $C_{l}^{(\sigma,R)}(t)=\bra n[\tilde{D}_{l}^{(\sigma,R)}]^{-1}(t)\tilde{D}_{l}^{(\sigma,R)}\ket n=\bra n[U_{\text{F}}^{(b,\mathcal{A})\dagger}]^{t}[U_{\text{F},R}^{(b,\mathcal{A}),\sigma}]^{t}\ket n$~\cite{dyn_dualtwist}.
This features the ``braided drive'' $U_{\text{F},R}^{(b,\mathcal{A}),\sigma}=R^{\dagger}U_{\text{F}}^{(b,\mathcal{A}),\sigma}R=i\tilde{\gamma}_{1}U_{\text{F}}^{(b,\mathcal{A}),\sigma}$ (cf. Fig.~\ref{fig:braid}).
Now $U_{\text{F},R}^{(b,\mathcal{A}),\sigma\dagger}\tilde{\gamma}_{q}U_{\text{F},R}^{(b,\mathcal{A}),\sigma}=(-1)^{q}\tilde{\gamma}_{q}$ ($q=1,2$): there is an \emph{unpaired} $\pi$ and zero mode each, both exact. 
Such unpaired modes are absent in existing TCs.

$\tilde{\gamma}_{1}$ toggles both $\mathbb{Z}_2$ and $\tilde{\mathbb{Z}}_2$ symmetries, exactly, $\smash{\overline{W}}^{(b)}\tilde{\gamma}_{1}=-\tilde{\gamma}_{1}\smash{\overline{W}}^{(b)}$ ($b\in\{e,m\}$). %
Thus, $C_{l}^{(\sigma,R)}(t)$ vanishes for odd $t$, and is nonzero otherwise, for \emph{any} $\smash{\overline{W}}^{(b)}$ eigenstate $\ket n$, and the braided drive pumps $\mathbb{Z}_{2}$ charge, exactly, between the system and its BCs.

\textit{From TCs to FCs via twisted temporal BCs:} Finally we use SymTFT again to further unify BC twisting concepts. 
First note that, in 2+1D spacetime, inserting $W_{j}^{(b)}$~\cite{Wdetails} yields a timelike $b$-anyon worldline at $j\in B_{\text{dyn}}$ [there is no worldline at $B_{\text{ref}}$ since $b$ or $\sigma(b)$ is in $\mathcal{A}_{\text{ref}}$]. 
Taking time to be periodic, as befits Floquet systems, this gives a noncontractible $W_{j}^{(b)}$ loop wrapping the temporal circle. 
``Unfolding" the cylinder in Fig.~\ref{fig:setup} makes this loop appear with period $L$ in space.
$b$-twisting temporal BCs is the space-time dual of this: the time periodic insertion of noncontractible loops $\smash{\overline{W}}^{(b)}$. 
Applied to an untwisted drive $U_{\text{F}}^{(\bm{1},\mathcal{A})}$ this precisely yields $\smash{\overline{W}}^{(b)}U_{\text{F}}^{(\bm{1},\mathcal{A})}=U_{\text{F}}^{(b,\mathcal{A})}$.
A drive $U_{\text{F}}^{(b,\mathcal{A})}$ thus amounts to $b$-twisted temporal BCs.

What is the space-time dual of duality-twisted BCs?
Duality-twisted spatial BCs feature a timelike worldsheet extending from $D_{\text{dyn}}^{(\sigma)}$. 
The space-time dual is a spacelike worldsheet $D_{\text{2D}}^{(\sigma)}$, repeated time periodically.
This holds for any automorphism $\sigma$ of a $G$-TO. %
Automorphism $\sigma$ applied time periodically to anyon worldlines is precisely the topological content of FCs~\cite{Hastings2021dynamically,Vuillot21,DavydovaPRXQ23,kesselring2022anyon,vu2023measurementinduced,sullivan2023floquet,aasen2023measurement,ellison2023floquet,davydova2023quantum,dua2023engineering}; %
these are thus $G$-TO systems with duality-twisted temporal BCs.
This viewpoint also extends to FC boundaries; these are such that logical operators (i.e., with ends condensing at $B_{\text{ref}}$ and $B_{\text{dyn}}$) remain logical operators over time~\cite{aasen2023measurement,ellison2023floquet,davydova2023quantum,dua2023engineering}; this implies $\mathcal{A}_{\text{(ref)}}\to\sigma[\mathcal{A}_{\text{(ref)}}]$, as achieved by bending $D_{\text{2D}}^{(\sigma)}$ forward in time at $B_{\text{ref}}$ and $B_{\text{dyn}}$ and continuing it in the future. 
For a period-$k$ automorphism, $\sigma^{k}=1$, this gives period $k$-tupling at FC boundaries; 
for $\mathbb{Z}_{2}$ FCs, by $\sigma^{2}=1$ for $e\leftrightarrow m$, this is the recently noted period doubling~\cite{aasen2023measurement}.

\textit{Conclusion:} 
We have shown how to incorporate driven MBL systems into SymTFT. 
This not only gave new insights into, and revealed new forms of, driven phases, 
but, via dynamical BCs and the incorporation of non-Abelian defect braids, it also enriched SymTFT.

Our considerations hold for any finite Abelian $G$ and they revealed striking new phenomena already for $G=\mathbb{Z}_2$. These may be realized using many-qubit devices akin to those that demonstrated $\mathbb{Z}_2$ TCs~\cite{mi2022time}. 
Going beyond $G=\mathbb{Z}_2$ may yield even richer ways to explore new Floquet phases, e.g., due to automorphisms spanning a much richer set. 
Exploring this is just one of the many exciting avenues SymTFT out of equilibrium offers for future research.

This work was supported by a Winton and an EPSRC Studentship and by EPSRC grant EP/V062654/1. 
Our simulations used resources at the Cambridge Service for Data Driven Discovery operated by the University of Cambridge Research Computing Service (\href{www.csd3.cam.ac.uk}{www.csd3.cam.ac.uk}), provided by Dell EMC and Intel using EPSRC Tier-2 funding via grant EP/T022159/1, and STFC DiRAC funding  (\href{www.dirac.ac.uk}{www.dirac.ac.uk}).

%\bibliography{ref}

\begin{thebibliography}{107}
\bibitem [{SM()}]{SM}%
  \BibitemOpen
  \href@noop {} {}\bibinfo {note} {The Appendix has further details on
  $\mathcal{H}_\text{dyn}$ and its embedding into the 2D TO, including charges,
  twisted BCs, and spectral degeneracies, on the absolute stability of the
  $0\pi$PM, and on the exact $\pi$ and zero modes.}\BibitemShut {Stop}%
\bibitem [{Wde()}]{Wdetails}%
  \BibitemOpen
  \href@noop {} {}\bibinfo {note} {If $a\neq\bar{a}$ the string is oriented
  $i\to j$ and we abbreviate $W_{i\to j}^{(a)}\equiv W_{ij}^{(a)}$ with
  $W_{ji}^{(a)}=W_{ij}^{(\bar{a})}=W_{ij}^{(a)\dagger}$. We terminate the
  $\text{Rep}\,G$ component of $W_{ij}^{(a)}$ on sites $i$ and $j$ while the
  $G$ component on links $\tilde{i}=\langle i,i+1\rangle$ and $\tilde{j}$ of
  $B_\text{dyn}$. Unless noted otherwise, the same applies to
  $W^{(a)}_j$.}\BibitemShut {Stop}%
\bibitem [{h05()}]{h05fn}%
  \BibitemOpen
  \href@noop {} {}\bibinfo {note} {We compute
  $\langle\hat{C}_{ij}^{\prime}(t)\rangle$ using only those
  $C_{ij}^{\prime}(t)$ with
  $|C_{ij}^{\prime}(0)|\equiv|C_{ij}^{\prime}|>10^{-6}$.}\BibitemShut {Stop}%
\bibitem [{dyn()}]{dyn_dualtwist}%
  \BibitemOpen
  \href@noop {} {}\bibinfo {note} {Duality twisting, strictly speaking, is not
  invertible; $[\tilde{D}^{(\sigma)}_l]^{-1}(\ldots)\tilde{D}^{(\sigma)}_l$
  means duality twisting $(\ldots)$.}\BibitemShut {Stop}%
\end{thebibliography}

%

\appendix
\renewcommand\thefigure{\thesection\arabic{figure}}
\setcounter{figure}{0}    

\section*{Appendix}

\section{Hilbert space, charges, boundary conditions, degeneracies}
\label{app:HS_C_BC}

In the main text we have discussed how the Hilbert space $\mathcal{H}_\text{dyn}$ of the 1D system is embedded into the 2D topological order (TO) as the subspace with no anyons away from the boundary $B_\text{dyn}$. 
Here we discuss how $\mathcal{H}_\text{dyn}$ decomposes into symmetry sectors and is enlarged with twisted sectors. 
We also provide further discussion on twisted boundary conditions (BCs) and spectral degeneracies. 
Our approach bridges between the condensed-matter oriented  Refs.~\onlinecite{lichtman2020bulk,TH23}, with Ref.~\onlinecite{TH23} working on the infinite cylinder and considering twisted BCs, while Ref.~\onlinecite{lichtman2020bulk} working on a finite cylinder (thus having SymTFT access to certain spectral degeneracies) but without twisted BCs.  
Ref.~\onlinecite{gaiotto2021orbifold} establishes results analogous to those discussed in this Appendix from a high-energy theory perspective. 
See also Refs.~\onlinecite{ChatterjeeWen23,JiWen2020categorical,Bhardwaj_PtII,BhardwajcatLand,Bhardwaj1_1}.

For finite Abelian $G$, we have $\mathcal{H}_{\text{dyn}}=\mathcal{H}_{1}^{\otimes L}$, with  $\mathcal{H}_{1}=\text{span}(\{\ket g,g\in G\})$.
$\mathcal{H}_{\text{dyn}}$ splits into symmetry sectors as $\mathcal{H}_{\text{dyn}}=\oplus_{\alpha}\mathcal{H}_{\text{dyn}}^{(1,\alpha)}$, where subspaces $\mathcal{H}_{\text{dyn}}^{(1,\alpha)}$ are of states ``charged" as $\alpha\in\text{Rep}\,G$ under the $G$ symmetry, $\smash{\overline{W}}^{(g,1)}\ket{\psi}=\chi_{\alpha}(g)\ket{\psi}$, $\forall\ket{\psi}\in\mathcal{H}_{\text{dyn}}^{(1,\alpha)}$.
Starting with the $G$-symmetric subspace $\mathcal{H}_{\text{dyn}}^{(1,1)}$, with $\smash{\overline{W}}^{(g,1)}\ket{\psi}=\ket{\psi}$, $\forall g\in G,\forall\ket{\psi}\in\mathcal{H}_{\text{dyn}}^{(1,1)}$, by $\smash{\overline{W}}^{(g,1)}W_{k}^{(1,\alpha)}=\chi_{\alpha}(g)W_{k}^{(1,\alpha)}\smash{\overline{W}}^{(g,1)}$, we have $\mathcal{H}_{\text{dyn}}^{(1,\alpha)}=W_{k}^{(1,\alpha)}\mathcal{H}_{\text{dyn}}^{(1,1)}$ (regardless of the choice of endpoint $k$).
The operators $W_{k}^{(1,\alpha)}$ thus introduce charge $(1,\alpha)$ to $B_\text{dyn}$ and thus to the 1D system.

The space $\mathcal{H}_{\text{dyn}}^{(g,\alpha)}=W_{j}^{(g,1)}\mathcal{H}_{\text{dyn}}^{(1,\alpha)}=W_{j}^{(g,1)}W_{k}^{(1,\alpha)}\mathcal{H}_{\text{dyn}}^{(1,1)}$ is of states obeying $g$-twisted BCs.
In the main text we interpreted this via the ``$G$-Aharonov Bohm" effect; here we provide a more direct interpretation: 
Consider $\mathcal{T}_{L}$, the translation of $B_{\text{dyn}}$, and with it $j$ and $k$, by $L$ sites, keeping the rest of the cylinder fixed. 
This wraps $W_{j}^{(g,1)}W_{k}^{(1,\alpha)}\equiv W^{(a)}$ [with $a=(g,\alpha)$] around the cylinder and, since reconnecting the resulting $a$ string  (more precisely, ribbon) amounts to a braid, we have $\mathcal{T}_{L}W^{(a)}\mathcal{T}_{L}^{\dagger}=e^{i\theta_{aa}}W^{(a)}\smash{\overline{W}}^{(a)}$. 
Thus, by $\smash{\overline{W}}^{(a)}\ket{\psi}=\mathcal{T}_{L}^{\dagger}\ket{\psi}=\ket{\psi}$ for $\ket{\psi}\in\mathcal{H}_{\text{dyn}}^{(1,1)}$, we find that $\mathcal{T}_{L}\ket{\psi}=\chi_{\alpha}(g)\ket{\psi}$ for $\ket{\psi}\in\mathcal{H}_{\text{dyn}}^{(g,\alpha)}$: the BCs are twisted. 
Note that the link $\tilde{j}=\langle j,j\!+\!1\rangle$ where we insert the BC twist is a gauge choice: it can be moved to $\tilde{j}'$ upon conjugating the Hamiltonian (or Floquet unitary) by $W_{jj'}^{(g,1)}$. 
To consider all twisted sectors together, we may work in the enlarged Hilbert space 
$\overline{\mathcal{H}}_{\text{dyn}}=\oplus_{\alpha,g}\mathcal{H}_{\text{dyn}}^{(g,\alpha)}$. 
When different BCs are coupled, as is the case in the presence of $\tilde{G}$-breaking perturbations, we must work with $\overline{\mathcal{H}}_{\text{dyn}}$. 

Turning to degeneracies, first recall that a gapped, or more generally many-body localized (MBL), 1D system is characterized by a Lagrangian subgroup $\mathcal{A}$; the $W_{k,k+1}^{(a)}$ with $a\in\mathcal{A}$ form a maximal set of local, mutually commuting $G$-symmetric operators, defining the zero-correlation-length (or fixed-point, FP) limit of the gapped, or MBL, phase. 
The topological ground-state degeneracy on the cylinder, equal to $|\mathcal{A}\cap\mathcal{A}_{\text{ref}}|$, i.e., the number of distinct strings that can connect $B_\text{ref}$  and $B_\text{dyn}$ without creating anyons, while avoiding crossing domain wall $D^{(\sigma)}_\text{ref}$, yields the ground-state (or spectral, for MBL) degeneracy of the 1D system within a twisted sector $\oplus_{\alpha}\mathcal{H}_{\text{dyn}}^{(g,\alpha)}$. 
Strings $W_{j}^{(a)}$ ($a\in\mathcal{A}$, $a\notin\mathcal{A}_{\text{ref}}$) that can similarly connect $B_\text{ref}$  and $B_\text{dyn}$ while (part of them) crossing $D^{(\sigma)}_\text{ref}$ yield degeneracies between twisted sectors. 
In the Floquet setting, both types of degeneracies may split into topological spectral multiplets (such as $\pi$ spectral pairing as discussed for $G=\mathbb{Z}_2$), with structure set by the drive's $\smash{\overline{W}}^{(b)}$.

\setcounter{figure}{0}    
\section{Absolute stability in the $0\pi$PM}
\label{app:0piPM_abs_stab}
As noted in the main text, the $\tilde{\mathbb{Z}}_{2}$ TC inherits the $\pi$SG's absolute stability~\cite{KeyserlingkKhemaniSondhi2016stability}. 
Here we study this, focusing on the robustness against $\tilde{\mathbb{Z}}_{2}$ symmetry breaking, and illustrating it on the spectral $\pi$-pairing in the $0\pi$PM phase. 

To study this, we consider $U_{\text{F}}^{(e,\langle m\rangle)}=\smash{\overline{W}}^{(e)}e^{-iH_1}e^{-iH_0}$ with $H_{0}=\sum_{j}J_{j}X_j+J_{j}^{(\text{int})}X_jX_{j+1}+\tilde{h}\tilde{X}_{1}$ and 
$H_{1}=\sum_{j}h_{j}W_{j,j+1}^{(e)}+h_{j}^{(\text{int})}W_{j,j+2}^{(e)}$.
With only $\tilde{X}_{1}$ as link operator in $H_0$, it is sufficient to just gauge its link, i.e. to take $W_{j,j+1}^{(e)}=Z_j Z_{j+1}$ for $j\neq 1$ and $W_{1,2}^{(e)}=Z_1\tilde{Z}_{1}Z_2$, and $W^{(e)}_L=\tilde{Z}_{1}$. 
Since $\smash{\overline{W}}^{(e)}\tilde{X}_{1}=-\tilde{X}_{1}\smash{\overline{W}}^{(e)}$, the $\tilde{h}\tilde{X}_{1}$ term breaks $\tilde{\mathbb{Z}}_{2}$ symmetry; it couples the system's untwisted, i.e., periodic boundary condition (PBC), and $m$-twisted, i.e., anti-periodic boundary condition (APBC), sectors.

We take $J_j$ distributed uniformly within $[\bar{J}/2,3\bar{J}/2]$ and $h_j$  within $[-\lambda\bar{J},\lambda\bar{J}]$ (with $\lambda\ll 1$), $\tilde{h}=\lambda\bar{J}$, and take  $J_{j}^{(\text{int})}=0.1\bar{J}$, $\bar{J}=3\pi/8$ for concreteness. 
The $\pi$-pairing is exact for $\lambda=0$ and is expected to receive corrections $\sim\lambda^L$ for small $\lambda$ because it requires at least order $L$ in perturbation theory for the $W^{(e)}_{j,j+1}$ to combine to $\smash{\overline{W}}^{(e)}$, thus contributing to a splitting between PBC and APBC, while $\tilde{h}\tilde{X}_{1}$, by the $\pi$ pairing not being a degeneracy, can contribute only in even orders of $\tilde{h}$ and only to shifts of the $\pi$ pair but not their splitting. 
[We set $h_{j}^{(\text{int})}=0$ for simplicity; with $h_{j}^{(\text{int})}\sim \lambda$ the corrections to exact $\pi$ pairing would be $\sim\lambda^{L/2}$.]
In contrast, the static PM's degeneracy between PBC and APBC sectors is split by $\tilde{h}\tilde{X}_{1}$ as $\sim\lambda$. 
Unlike this, the $\pi$ pairing in the  $0\pi$PM receives corrections exponentially small in $L$, hence the absolute stability against $\tilde{Z}_2$ symmetry breaking. 
(One can more formally prove absolute stability in the $0\pi$PM using local unitary reasoning similar to that in Ref.~\onlinecite{KeyserlingkKhemaniSondhi2016stability}.)

\begin{figure}[t]
 \includegraphics[width=0.9\columnwidth]{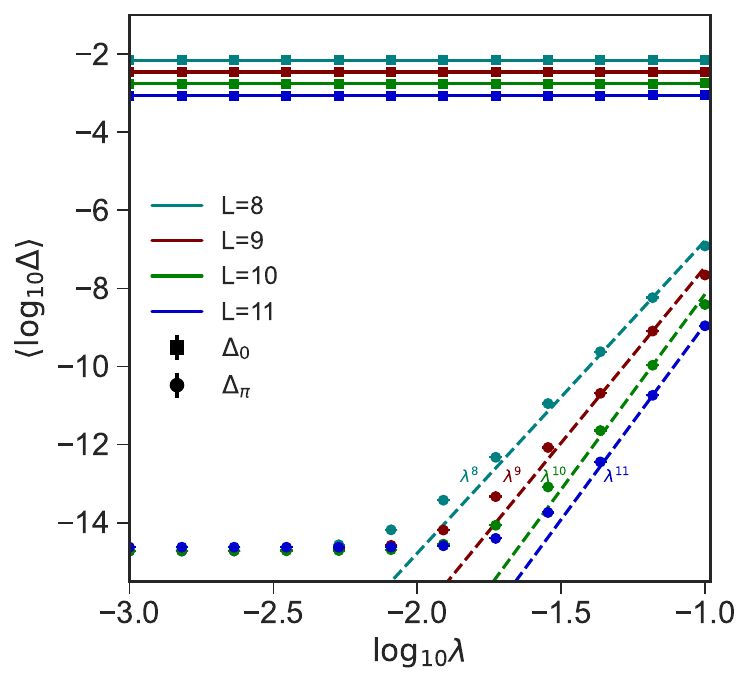}
 \caption{The log-average level spacing $\Delta_0$ and deviation $\Delta_\pi$ from $\pi$-pairing for the model in the text, where $\langle\ldots\rangle$ denotes averaging over all eigenstates and $1000$ disorder realizations. 
 Dashed lines show $\lambda^L$. 
 Error bars, at the standard error of the log-average, are imperceptible.
 The small-$\lambda$ saturation in  $\Delta_\pi$ is due to $\pi$ pairing becoming accurate to machine precision.
 }\label{fig:abs_stab}
\end{figure}

We next test absolute stability numerically. 
We define~\cite{KeyserlingkKhemaniSondhi2016stability} $\Delta_{0}^{i}=\varepsilon_{i+1}-\varepsilon_{i}$, with $\varepsilon_{i}$ the $i^{\text{th}}$ quasienergy of $U_{\text{F}}^{(e,\langle m\rangle)}$ and $\Delta_{\pi}^{i}=|\varepsilon_{i+M/2}-\varepsilon_{i}-\pi|$, where $M=2^{L+1}$ (the dimension of $\overline{\mathcal{H}}_{\text{dyn}}$ including both PBC and APBC sectors), i.e., $i+M/2$ is halfway across the spectrum from $i$. 
We expect the typical values $\langle\ldots\rangle_\text{typ}=\exp(\langle\log(\ldots)\rangle)$ to satisfy $\langle\Delta_{0}\rangle_\text{typ}\sim e^{-sL}$ (with $s\sim\ln2$ independent of $\lambda$) while $\langle\Delta_{\pi}\rangle_\text{typ}\sim\lambda^{L}$; we focus on $\Delta_{\pi}<\Delta_{0}$ to be able to identify $\pi$ spectral pairs in the many-body spectrum. 
Our numerical results are shown in Fig.~\ref{fig:abs_stab}. 

We focused on breaking $\tilde{\mathbb{Z}}_{2}$ symmetry on a single site for simplicity: this required only one extra qubit (for $\tilde{Z}_{1}$, $\tilde{X}_{1}$). 
With just one gauged link, we could not demonstrate the robustness against $\mathbb{Z}_2$ symmetry breaking since with a $\mathbb{Z}_2$-breaking perturbation $H_{z}=\sum_{j}h_{j}^{z}Z_{j}$ and $h_j^z\sim \lambda$, the $\pi$ pairing would receive corrections already at order $\lambda^3$ due to $Z_1 \tilde{Z}_{1} Z_2$, $Z_1$ and $Z_2$ being able to combine into $\smash{\overline{W}}^{(e)}=\tilde{Z}_{1}$. 
However, upon gauging all links, $W^{(e)}_{j,j+1}\to Z_j\tilde{Z}_{j}Z_{j+1}$ and thus $\smash{\overline{W}}^{(e)}=\prod_{j=1}^L \tilde{Z}_{j}$, we would find that the corrections to $\pi$ pairing scale as $\lambda^L$ even with $H_Z$ present. 
Testing this would require doubling not the Hilbert space, but the number of qubits, and is thus numerically much more challenging.

\setcounter{figure}{0}    
\section{Exact $\pi$ and zero modes in free-fermion Ising drives}
\label{app:Ising_free}

Here we provide further details on the $\pi$ and zero modes in (braided) drives arising from duality-twisted BCs. 
In particular, to illuminate their exact nature and persistence, even as criticality is approached, we illustrate them on a $\mathbb{Z}_2$-symmetric (i.e., Ising) system admitting a free-fermion description upon Jordan-Wigner (JW) transformation. 
We emphasize, however, that the exactness and persistence of these modes hold more generally, including with interactions and for any finite Abelian symmetry $G$.
Our discussion in this Appendix builds on Ref.~\onlinecite{aasen2016topological}'s discussion of duality-twisted BCs in the Ising model, and complements  Refs.~\onlinecite{tan2022topological,Mitra23,samanta2023isolated} which studied duality twisting the Floquet SG and PM, i.e., the ``static" phases (without $\pi$ modes) of the Floquet phase diagram.

\subsection{Duality twisted local Ising Hamiltonians and exact zero modes}

We start with inspecting the $\mathbb{Z}_2$-symmetric algebra after duality-twisting BCs. 
The local generators are
\begin{equation}
Z_{l-1}Y_{l},\  X_{j},\ Z_{j-1}Z_{j},\ (j\neq l),
\label{eq:Ising_loc}
\end{equation}
and the nonlocal generators are
$
X_{l},\ Z_{l-1}Z_{l}
$.
We also include the link operator $\tilde{Z}_{l-1}$, labeling BCs, updating $Z_{l-1}Z_{l}\to Z_{l-1}\tilde{Z}_{l-1}Z_{l}$, hence $Z_{l-1}Y_{l}\to Z_{l-1}\tilde{Z}_{l-1}Y_{l}$. 
In what follows, we set $l=1$, without loss of generality, thus $Z_{l-1}\tilde{Z}_{l-1}Y_{l}=Z_{L}\tilde{Z}_{L}Y_{1}$ for an $L$-site system.

We next focus on non-interacting Ising systems, i.e., whose Hamiltonians involve terms from Eq.~\eqref{eq:Ising_loc}, but not their products. 
In this case, it is useful to introduce $2L$ Majorana operators $c_{i}$, and an auxiliary
pair $\Gamma_{1,2}$ for $\tilde{Z}_{L}$, via JW transformation.
We have 
\begin{multline}
X_{j}=ic_{2j-1}c_{2j}\ \forall j,\ Z_{j}Z_{j+1}=ic_{2j}c_{2j+1}\ (j\neq L),\\ Z_{1}Z_{L}=\prod_{j=1}^{L-1}Z_{j}Z_{j+1}=\prod_{j=1}^{L-1}ic_{2j}c_{2j+1},\ \tilde{Z}_{L}=i\Gamma_{1}\Gamma_{2},
\label{eq:JW1}
\end{multline}
and 
\begin{multline}
P\equiv\prod_{j=1}^{L}X_{j}=\prod_{j=1}^{L}(ic_{2j-1}c_{2j})\ \Rightarrow\ Z_{L}Z_{1}=-iPc_{2L}c_{1},\\ \!\!\!Z_{L}Y_{1}=P(ic_{2L}c_{2}),\ Z_{L}\tilde{Z}_{L}Y_{1}=P(i\Gamma_{1}\Gamma_{2})(ic_{2L}c_{2}).
\label{eq:c1_out}
\end{multline}
Eq.~\eqref{eq:c1_out} shows that $c_{1}$ enters local couplings only via $P$ in $Z_{N}Y_{1}$ and thus $c_{1}$ commutes with all terms, except $\{c_{1},Z_{N}\tilde{Z}_{L}Y_{1}\}=0$. 
Accompanying it with, say, $\Gamma_{1}$,  effecting a BC twist, the operator $\tilde{\gamma}_{1}=ic_{1}\Gamma_{1}$ commutes with all local terms, hence it is an exact zero mode. 
[Ref.~\onlinecite{aasen2016topological} emphasizes that such Ising Hamiltonians, with closed boundaries, do \emph{not} have zero modes despite their exactly degenerate spectrum. 
This however pertains to a given twisted sector and is hence consistent with our finding since $\tilde{\gamma}_{1}$ connects distinct twisted sectors. 
The exact degeneracy $\tilde{\gamma}_{1}$ implies between these sectors holds also with interactions, and generalizes to other finite Abelian groups; in contrast, the exact degeneracy (i.e., without finite-size corrections) within twisted sectors appears not to generalize beyond $G=\mathbb{Z}_2$.]

Since all local $\mathbb{Z}_{2}$-symmetric (and BC preserving) terms commute with $P$ and $\tilde{Z}_{L}$, we can consider each of the, $P\tilde{Z}_{L}=p\tilde{z}\equiv s=\pm1$ sectors separately (here $p,\tilde{z}$ are the eigenvalues; $s$ is their product);
within these, a Hamiltonian $H_{s}$ that is a linear combination of the local terms [Eq.~\eqref{eq:Ising_loc}] is quadratic in Majorana operators $c_j$. 
Now, since $s$ is a number, $c_{1}$ is absent from $H_{s}$: $\gamma_{1}\equiv c_{1}$ is an exact zero mode of $H_{s}$ (but not of $H$). 
Since $H_{s}$ is a quadratic fermion Hamiltonian, if it has one exact zero mode, it must have another.
This is indeed the case: from
\[
H_{s}=i\sum_{j,k\neq1}h_{jk}^{(s)}c_{j}c_{k},
\]
the spectrum is governed by $h_{jk}^{(s)}$; this is $(2L-1)\times(2L-1)$ antisymmetric (and real) hence always has a zero eigenvalue. 
The corresponding eigenvector $v^{(s)}$ (also real) supplies the other zero mode $\gamma_{2}^{(s)}=\sum_{j=2}^{2L}v_{j}^{(s)}c_{j}$.

These two exact zero modes of $H_{s}$ are there regardless of the values of the couplings, including at criticality. 
One of them is always $\gamma_{1}=c_1$; it is perfectly localized. 
The localization of $\gamma_{2}^{(s)}$ depends on microscopic details; in terms of the MBL localization length $\xi$ we expect it to have a profile concentrated within $\sim\xi$ of the site nearest to twist $T_2^{(\sigma)}$ that remained near $B_\text{dyn}$ (in terms of SymTFT) after duality twisting BCs. 
In terms of the 1D system, the position of $T_2^{(\sigma)}$ marks an interface between opposite  eigenstate orders, i.e., between $\mathcal{A}=\langle e\rangle$ (SG) and $\mathcal{A}=\langle \sigma(e)\rangle=\langle m\rangle$ (PM). This translates to an interface between topologically distinct regions in the fermion language, binding a Majorana zero mode.

The striking feature of duality-twisted BCs is that $\gamma_{1}$ and $\gamma_{2}^{(s)}$ are exact, without the spectrum incurring any splitting from the zero modes' non-infinite spatial separation. 
Indeed, the zero modes persist even if they are right next to each other in 1D, such as for $D_\text{dyn}^{(\sigma)}$ extending just along a single link along $B_\text{dyn}$ (as in the main text).
The SymTFT intuition for this is that, even in this case, the zero modes are  ``infinitely far" apart, with $\tilde{\gamma}_1$ being attached to $T_1^{(\sigma)}$ deep in the 2D bulk while $\tilde{\gamma}_2$ being at  $T_1^{(\sigma)}$ near $B_\text{dyn}$.

How do these zero modes enter the full Hamiltonian $H$? 
To study this, we split $v^{(s)}$ into even and odd parts in $s$, i.e., $v^{(s)}=\frac{1}{2}[v^{(+)}+v^{(-)}]+\frac{1}{2}s[v^{(+)}-v^{(-)}]\equiv v^{(\text{e})}+sv^{(\text{o})}$.
Introduce the operator 
\begin{equation}
\gamma_{2}=\sum_{j=2}^{2L}(v^{(\text{e})}_{j}+Sv^{(\text{o})}_{j})c_{j},
\label{eq:gamma_2MB}
\end{equation}
where $S=P\tilde{Z}_{L}$. 
This is not fermion linear, not local, and not even Hermitian. 
However, the combination $\Psi_{0}=i\gamma_{2}\gamma_{1}$ has the following familiar features: 
it is Hermitian, $\Psi_{0}^{2}=\id$ and it exactly commutes with $H$. 

Noting that $\Gamma_{1}$ and $\gamma_1=c_{1}$ are each absent from $H_{s}$ and anticommute with $c_{j\neq1}$ and with $Z_{L}Y_{1}\tilde{Z}_{L}$, we can replace $\gamma_1\to \Gamma_{1}$ in $\Psi_0$ to find that $\tilde{\gamma}_{2}=i\gamma_{2}\Gamma_{1}$ is also Hermitian, exactly commutes
with $H$, and satisfies $\tilde{\gamma}_{2}^{2}=\id$, $\{\tilde{\gamma}_{1},\tilde{\gamma}_{2}\}=0$. 
Recalling $\tilde{\gamma}_{1}=i\gamma_{1}\Gamma_{1}$, we have $\Psi_{0}=i\tilde{\gamma}_{2}\tilde{\gamma}_{1}$.
Thus, in $\Psi_0$ we recognize the nonlocal integral of motion $\mathcal{W}_{\sigma}^{(a)}$ described in the main text. 

\subsection{$\pi$ and zero modes in (braided) Ising drives}

These considerations naturally extend to the Floquet context. 
We write the duality-twisted Floquet unitary as $U_{\text{F}}^{(b,\mathcal{A}),\sigma}=\smash{\overline{W}}^{(b)}U_{\text{F}}^{(\bm{1},\mathcal{A}),\sigma}$. 
The drive factor $U_{\text{F}}^{(\bm{1},\mathcal{A}),\sigma}$ for the twist configuration  in the main text [i.e., with $D_\text{dyn}^{(\sigma)}$ extending just from $l-1/4$ to $l+3/4$ along the $B_\text{dyn}$ direction], and for $\mathcal{A}=\langle e\rangle$ for concreteness, is
\[
U_{\text{F}}^{(\bm{1},\langle e\rangle),\sigma} = e^{-i\sum_{j\neq1}h_{j} X_{j}}e^{-i J_{1}Z_{L}\tilde{Z}_{L}Y_{1}}e^{-i\sum_{j\neq1}\kappa_{j} Z_{j-1}Z_{j}},
\]
where $\kappa_2=h^\prime_j$ and $\kappa_{2<j\leq L}=J_j$, where the $J_j$ couplings set the dominant energy scale (hence the eigenstate order) and the $h^{(\prime)}_j$ couplings are perturbations. 
[We split $U_{\text{F}}^{(\bm{1},\langle e\rangle),\sigma}$ into three factors, each with mutually commuting terms; however the precise splitting conventions are immaterial for the exact zero and $\pi$ modes: the key is that $\tilde{\gamma}_1$ commutes with all local generators, Eq.~\eqref{eq:Ising_loc}.]

By it involving only terms from Eqs.~\eqref{eq:JW1} and \eqref{eq:c1_out}, we can describe
$U_{\text{F}}^{(\bm{1},\mathcal{A}),\sigma}$ using free fermions %
in each of the four $S=P,\tilde{Z}_{L}$ eigenspaces (with eigenvalues $p$, $\tilde{z}$ and $s=p\tilde{z}$ as above). 
Turning to $U_{\text{F}}^{(b,\mathcal{A}),\sigma}$, the line operator $\smash{\overline{W}}^{(b)}$
does not cause difficulties, since $\smash{\overline{W}}^{(b)}=P^{r_{m}}\tilde{Z}_{L}^{r_{e}}$ with $r_{m},r_{e}\in\{0,1\}$.
For each $p,\tilde{z}$, the unitary $U_{\text{F}}^{(b,\mathcal{A}),\sigma}$ thus acts via quadratic fermion gates; we denote the corresponding operator by $U_{\text{F},p,\tilde{z}}^{(b,\mathcal{A}),\sigma}=p^{r_{m}}\tilde{z}^{r_{e}}U_{\text{F},s}^{(\bm{1},\mathcal{A}),\sigma}$. 

As before, $c_{1}$ is absent, hence $\gamma_1=c_1$ is an exact zero mode of $U_{\text{F},s}^{(\bm{1},\mathcal{A}),\sigma}$. 
We find its pair using $e^{-i\sum_{j\neq1}h_{j}X_{j}}=e^{\sum_{ij}q_{ij}^{X}c_{i}c_{j}}$, $e^{-i\sum_{j\neq1}\kappa_{j}Z_{j-1}Z_{j}}=e^{\sum_{ij}q_{ij}^{Z}c_{i}c_{j}}$, 
$e^{-iJ_{1}Z_{L}\tilde{Z}_{L}Y_{1}}\to e^{\sum_{ij}q_{ij}^{1,s}c_{i}c_{j}}$
with $(2L-1)\times(2L-1)$ antisymmetric matrices $q_{ij}^{Z},\ q_{ij}^{X},\ q_{ij}^{1,s}$.
Using the correspondence between an operator $\hat{Q}=e^{\sum_{jk} q_{jk} c_j c_k}$ and matrix $Q=e^{4q}$ in  $\hat{Q} c_n \hat{Q}^{-1}=\sum _m Q_{mn}c_m$~\cite{kitaev2001unpaired}, the sought pair is $\gamma_{2}^{(s)}=\sum_{j=2}^{2L}v_{j}^{(s)}c_{j}$, with $v^{(s)}$  now satisfying $e^{4q^{X}}e^{4q^{1,s}}e^{4q^{Z}}v^{(s)}=v^{(s)}$. 

\begin{figure}[b]
 \includegraphics[width=0.9\columnwidth]{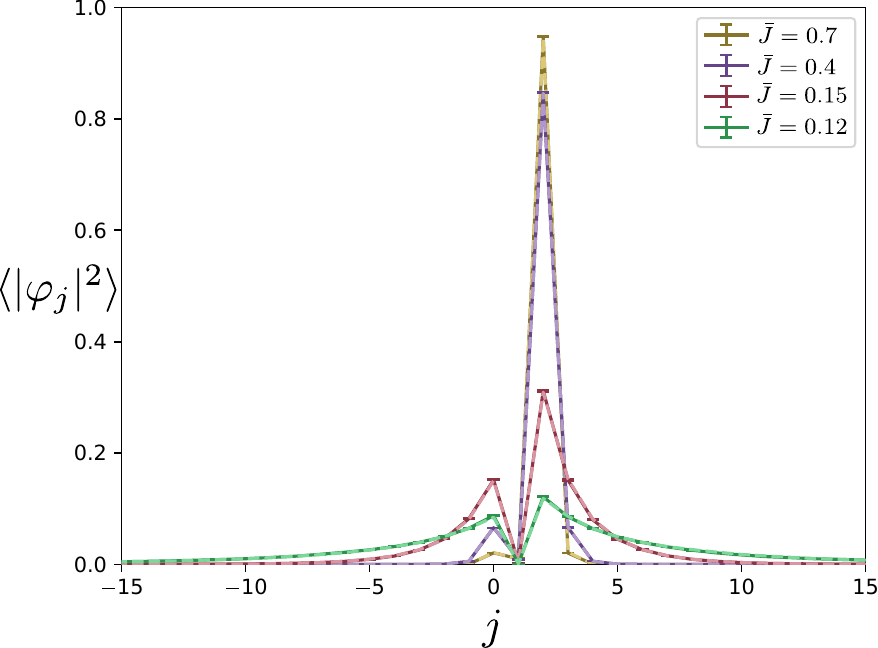}
 \caption{Numerical results on the disorder average of $|\varphi_j|^2=|v^{(s)}_{2j-1}|^2+|v^{(s)}_{2j}|^2$ for the Majorana wavefunction $v^{(s)}$. Solid lines show $s=1$; dashed lines are $s=-1$ (the difference is almost imperceptible).
 We used $L=100$ (i.e., $200$ Majorana sites), sampled $J_j$ uniformly from $[\bar{J}/2,3\bar{J}/2]$ and $h_j$ from $[\bar{h}/2, 3\bar{h}/2]$, with $\bar{h}=0.1$. 
 The plots for $\bar{J}=0.7,\,0.4$ used $2000$ and $5000$ realizations, respectively, while those for  $\bar{J}=0.15,\,0.12$ each used $10000$ realizations. 
 Error bars show the standard error of the mean. 
   }\label{fig:zm_profile}
\end{figure}

By $q_{ij}^{Z},\ q_{ij}^{X},\ q_{ij}^{1,s}$ being odd dimensional antisymmetric, such a $v^{(s)}$ always exists and, sufficiently away from criticality (i.e., from $\bar{J}=\bar{h}$ for the typical values $\bar{J}$ of $J_i$ and $\bar{h}$ of $h_j$~\cite{KhemaniLazaridesMoessnerSondhi2016TC,ElseTC2016,KeyserlingkSondhi2016floquetSPT,KeyserlingkSondhi2016floquetSSB,KeyserlingkKhemaniSondhi2016stability,KhemaniKeyserlingkSondhi2017RepTh,time_crystals_review,Berdanier_2018}) is unique. 
The value of $\bar{J}/\bar{h}$, however, sets the spatial profile of $v^{(s)}$, making it increasingly delocalized upon approaching $\bar{J}=\bar{h}$. 
We illustrate this via numerical simulations, shown in Fig.~\ref{fig:zm_profile}.

From $\gamma_{2}^{(s)}$ one gets the many-body operator $\gamma_2$ as in Eq.~\eqref{eq:gamma_2MB}. 
For $U_{\text{F}}^{(\bm{1},\mathcal{A}),\sigma}$, the operators $\tilde{\gamma}_{1}=i\gamma_{1}\Gamma_{1}$ and $\tilde{\gamma}_{2}=i\gamma_{2}\Gamma_{1}$ are again exact zero modes. 
For $U_{\text{F}}^{(b,\mathcal{A}),\sigma}$ (with $\mathcal{A}=\langle e\rangle$, $b=m$ or vice versa), by $\smash{\overline{W}}^{(b)}\tilde{\gamma}_{j}=-\tilde{\gamma}_{j}\smash{\overline{W}}^{(b)}$ ($j=1,2$), these are both exact $\pi$ modes. 
Finally, for the braided drive $U_{\text{F},R}^{(b,\mathcal{A}),\sigma}=i\tilde{\gamma}_{1} U_{\text{F}}^{(b,\mathcal{A}),\sigma}$, the operator $\tilde{\gamma}_{1}$ is an exact $\pi$ mode, while, by $\tilde{\gamma}_{1}\tilde{\gamma}_{2}=-\tilde{\gamma}_{2}\tilde{\gamma}_{1}$, the operator  $\tilde{\gamma}_{2}$ is an exact zero mode.
Thus we have an unpaired $\pi$ and zero mode each, both exact.

\end{document}